\documentclass[prl,twocolumn,showpacs,floatfix]{revtex4}

\usepackage{graphicx}
\usepackage{array}
\usepackage{amsmath,amsfonts,amssymb}
\usepackage[ansinew]{inputenc}
\usepackage{color}

\newcommand{\Figref}[1]{Fig.~\ref{#1}}

\begin{document}

\title{Oligothiophene nano-rings as electron resonators for whispering gallery modes}

\author
{Gaël Reecht,$^1$ Hervé Bulou,$^1$ Fabrice Scheurer,$^1$ Virginie Speisser,$^1$ Bernard Carrière,$^1$ Fabrice Mathevet,$^2$ Guillaume Schull,$^{1}$ }

\altaffiliation{ guillaume.schull@ipcms.unistra.fr}
\affiliation{$^1$ IPCMS de Strasbourg, UMR 7504 (CNRS -- Universit\'e de Strasbourg), 67034 Strasbourg, France}
\affiliation{$^2$ Laboratoire de Chimie des Polym\`eres, UMR 7610 (CNRS -- Universit\'e Pierre et Marie Curie), 75252 Paris, France}

\begin{abstract}
Structural and electronic properties of oligothiophene nano-wires and rings synthesized on a Au(111) surface are investigated by scanning tunneling microscopy. The spectroscopic data of the linear and cyclic oligomers show remarkable differences which, to a first approximation, can be accounted by considering electronic states confinement to one-dimensional (1D) boxes having respectively fixed and periodic boundary conditions. A more detailed analysis shows that polythiophene must be treated as a ribbon (i.e. having an effective width) rather than a purely 1D structure. A fascinating consequence is that the molecular nano-rings act as whispering gallery mode resonators for electrons, opening the way for new applications in quantum-electronics.   
\end{abstract}

\date{\today}


\maketitle

Whispering galleries, such as the dome of St Paul's Cathedral in London \cite{rayleigh1910}, are intriguing structures conveying waves on a curved path. For closed-loop galleries, wave resonances appear when an integer number of wavelengths matches the perimeter of the resonator. These whispering gallery modes (WGM) exist for all types of waves and numerous applications emerged, from guided propagation of radio waves in the ionosphere \cite{budden1962,Carrara1970} to confinement of light in optical WGM resonators \cite{Liu1997,Vahala2003,ilchenko2006,Vollmer2008,Righini2011}. For WGM to exist, the coherence length of the waves must exceed the perimeter of the resonator, and the walls must efficiently reflect the waves.  Circular walls of several meters, present in many monuments around the world, fulfil these conditions for acoustic waves \cite{rayleigh1910}. For optical waves, micrometer resonators of various shapes supporting WGM were designed \cite{Vahala2003}. In this case, the reflections at the outer wall rely on the total internal reflection at the medium--air boundary. More recently, whispering galleries were built for plasmons \cite{Cho2011} and neutrons \cite{Nesvizhevsky2010} and WGM predicted for anti-hydrogen atoms \cite{Voronin2012}. While future electronic devices may highly benefit from this concept in charge transport, no whispering gallery effect for electrons has been evidenced so far. For electrons, a WGM-resonator must be a nanoscale closed-curved conductive structure, reflecting electronic waves at its boundaries. One-dimensional (1D) extended molecules have recently attracted a tremendous interest because of potential use as electrical nano-wires \cite{HoChoi2008,Lafferentz2009,schull2009}. The electronic structure of these oligomers can be approximated by a (nearly) free electron gas confined to a 1D box \cite{Repp2010,Wang2011}. Here we use the intrinsic flexibility of a conjugated molecular wire, oligothiophene, to confine these electronic states in nano-rings. The electronic properties of these rings are energetically and spatially resolved using scanning tunneling microscopy. This study reveals molecular resonances with homogeneous ring-shape spatial repartition whose diameter increases with energy, in agreement with the formalism describing WGM \cite{Pluchon2012}.

\begin{figure}
  \includegraphics[width=.99\linewidth]{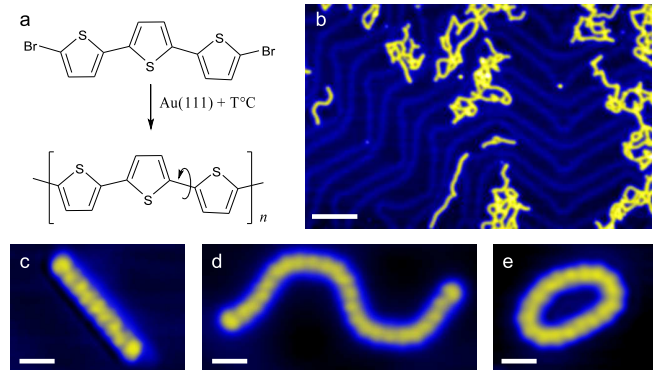}
  \caption{(Color online) (a) Sketch of the on-surface polymerization reaction of DBrTT on Au(111). (b) Large scale topographic STM image ($I$ = 100 pA and $V_{sample}$ = 0.1 V) and close-up views (c, d, e) of various oligothiophene structures resulting from the polymerization reaction. The white scale bars correspond 5 nm (b) and 1 nm (c, d, e).}
\label{fig1} 
\end{figure}

\noindent
The experiments were performed with an Omicron STM operated at 4.6\,K in ultrahigh vacuum. Au(111) samples and chemically etched W tips were prepared by Ar$^+$ bombardment and annealing. 5,5''-Dibromo-2,2':5',2''-terthiophene (DBrTT) molecules (\Figref{fig1}a) purchased from Sigma-Aldrich were evaporated from a Ta crucible. The Au(111) samples were heated at 400\,K during the deposition and at 550\,K for 30 minutes after to induce the on-surface polymerisation \cite{Grill2007} of the molecules (\Figref{fig1}a).
The data shown correspond to a coverage of approximately 0.1 monolayer \cite{note1}. Differential conductance spectra (maps) were acquired using lock-in detection with a modulation frequency of 740 Hz and a root-mean-square modulation amplitude of 10 (25) mV. All spectroscopic data were recorded at \textit{constant height} to prevent artefacts linked to the trajectory of the STM tip.

\noindent
A scanning tunneling microcopy (STM) image of oligothiophene wires is shown in figure \ref{fig1}b. High-resolved images of individual strands (\Figref{fig1}c, d, e) reveal intra-molecular structures spaced by $\approx$ 0.38\,nm, matching the distance measured between thiophene bases in oligothiophene \cite{Kasai2002,Grevin2003,bocheux2011}. In contrast to poly(\textit{p}-phenylene) \cite{Lipton2009,Wang2011} which generally adopts a linear conformation, oligothiophene has the intrinsic ability to fold \cite{Kasai2002,Grevin2003,bocheux2011} because of the monomer's pentagonal geometry. Successive thiophene units can adopt two conformations: one where the sulfur atoms are opposite to each other and a second where they are on the same side of the wire. While the former leads to linear strands (\Figref{fig1}c) the latter leads to bended molecular wires (\Figref{fig1}d). Close-ended cyclic oligomers\cite{Kromer2000} of different diameters also naturally form on the surface (\Figref{fig1}e). We will see below that, despite these folds, the electronic conjugation is preserved along the wires. Consequently, oligothiophene is an ideal candidate for electrical atomic-scale wires requiring high flexibility. On-surface polymerisation which was never used so far to create thiophene-based polymers allows synthesizing and studying \textit{pure} thiophene macro-cycles for the first time here. This method avoids solubilizing side-groups, mandatory for syntheses in solution \cite{Kromer2000}, which usually affect the conformations and electronic properties of oligothiophene.\\
\begin{figure}
  \includegraphics[width=.99\linewidth]{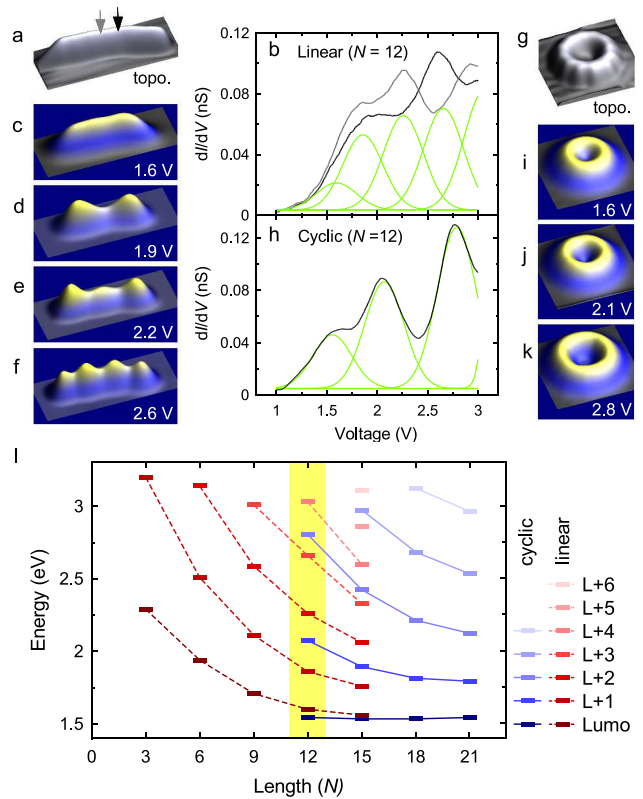}
  \caption{(Color online) (a) Topographic STM image ($I$ = 100 pA and $V_{sample}$ = 0.1 V, 6.6 $\times$ 2.4 nm$^2$) and (b) constant height differential conductance spectra (set point: $I$ = 5 pA and $V_{sample}$ = 1 V) of a linear-[$12$]-thiophene. The grey and black spectra correspond to two positions of the tip on top of the wire (see grey and black arrows in (a)). Green lines are Gaussian function fits. (c) to (f) are constant height conductance maps acquired at voltages corresponding to maxima in (b). The same data were acquired for a cyclo-[$12$]-thiophene: (g) topographic STM image (2.8 $\times$ 2.8 nm$^2$), (h) constant height differential conductance spectra acquired on top of the wire, and (i) to (k) constant height conductance maps acquired at voltages corresponding to maxima in (h). No resonances were resolved at negative sample voltages. (l) Molecular state energies as a function of oligomer lengths for linear- (red) and cyclo- (blue) [$N$]-thiophene ([$12$]-thiophene wires are underlined in yellow).}
\label{fig2}
\end{figure}

\noindent
To see how the closed-loop architecture of the rings impacts their electronic properties, a comparison with open-ended wires is mandatory. Figures \ref{fig2}a and g respectively display STM topographic images of a linear and a cyclic $N$ = 12 oligothiophene (with $N$ the number of thiophene bases). Differential conductance (d$I$/d$V$) spectra (\Figref{fig2}b and h) identify the lowest unoccupied molecular orbitals (LUMOs) for the linear and cyclic structures. While the intensities of the conductance peaks vary as a function of tip position along the linear wires (e.g. \Figref{fig2}b), the spectra remain unchanged along the cyclic structure. The number and the energy of the resonances are different for the two oligomers. Fitting the data with Gaussian distributions (green lines in \Figref{fig2}b and h) reveals that the resonances have similar widths ($\approx$ 0.4 eV). Details of the fitting procedure are available in \cite{supmat}. Figures \ref{fig2}c to f (i to k) show differential conductance maps acquired at constant height for the linear (cyclic) wire at voltages corresponding to maxima in the spectrum of \Figref{fig2}b (h). For the LUMO+$N$ orbital, the conductance maps reveal $N$+1 bright lobes along the linear wire, explaining the changes in the d$I$/d$V$ spectra with tip position. This behaviour can be understood from the confinement of states to a one-dimensional box as for oligothiophene adsorbed on NaCl \cite{Repp2010}. For the cyclo-[$12$]-thiophene, the conductance maps  reveal a different behaviour. For the three considered orbitals, a uniform ring (i.e. without nodes) is observed, whose diameter strikingly increases with the energy of the state.
Spectroscopic data as a function of $N$ (Fig.\,\ref{fig2}l) reveal a downward shift in energy and a reduction of the energy gaps between resonances when $N$ increases for both linear and cyclic oligomers. This is the signature of electron delocalisation \cite{Telesca2001}, definitely evidencing the conjugated nature of the oligomers.

\begin{figure}
  \includegraphics[width=.99\linewidth]{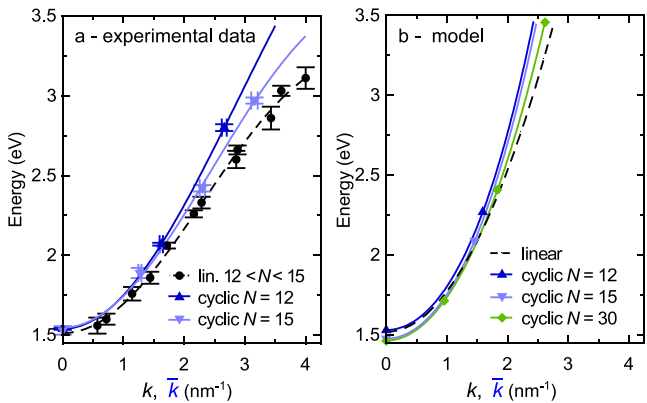}
  \caption{(Color online) Experimental (a) and calculated (b) dispersion data for linear and cyclic oligomers. The connecting lines correspond to fits with a 1D tight-binding expression. In (b) the parabolic dispersion of an infinite wire is represented (dashed lines). Details about the fits and the error bars are provided in \cite{supmat}.}
\label{fig3}
\end{figure}

\noindent
Figure \ref{fig2}l also confirms that the orbitals of linear and cyclic oligomers of equal length do not match. This apparent discrepancy finds its origin in the close-ended nature of the cyclo-[$N$]-thiophene resonances which must be treated with periodic boundary conditions \cite{Bednarz2004}. This has two important consequences: (i) States exist only when a integer multiple of the wavelength matches the length of the cyclic structures (half-wavelength for the linear wires). Consequently, the wavenumbers of linear- and cyclo-$[N$]-thiophenes are given respectively by $k=l\frac{\pi}{L}$ and $\bar k =l\frac{2\pi}{L}$, where $l$ is the quantum state number, and $L$ the delocalisation length \cite{supmat}; (ii) Contrary to open-ended wires where wave functions vanish at the boundaries, the $l$ = 0 resonance exists. The energy of this state is invariant with $N$, as experimentally observed (i.e. LUMO of the cyclic oligomers in \Figref{fig2}l). Figure \ref{fig3}a displays the deduced dispersion data for linear- and cyclo-[$N$]-thiophene for $N$ = 12 and 15. The lines in \Figref{fig3}a correspond to fits with a 1D tight-binding expression (see details in \cite{supmat}). The overall shape of the dispersion curves is very similar for the two conformations, in agreement with the fact that the electronic properties of conjugated oligomers essentially derive from their length \cite{Telesca2001}.

\noindent
A close inspection of \Figref{fig3}a reveals small, yet measurable, differences between the dispersion curves of the linear and cyclic structures.
Unlike linear chains where the dispersion is length-independent, the dispersion gets stronger with decreasing $N$ for cyclo-[$N$]-thiophenes.
This behaviour, as well as the increased radius of the orbitals of the cyclic oligomers with energy (\Figref{fig2}i to k), can be accounted by considering polythiophene as a 2D ribbon rather than a 1D wire. 
In this framework, the wavefunction of an electron of effective mass $m^{*}$ moving in an annular potential $V(r)$ is given in polar coordinates by $\psi(\vec{r})={R(r)}{Y(\theta)}$ where $R(r)=\frac{u(r)}{\sqrt(r)}$ and $Y(\theta)$ are respectively the radial and angular parts, and satisfy:

\begin{equation}
\label{eq:radial}
 -\frac{\hbar^2}{2m^{*}}\frac{d^2u\left(r\right)}{dr^2}+\left[V\left(r\right)+\frac{\hbar^2}{2m^{*}}\frac{1}{r^2}\left(l^2-\frac{1}{4}\right)\right]u\left(r\right)=\epsilon u\left(r\right),
\end{equation}

\begin{equation}
\frac{d^2Y(\theta)}{d\theta^2}=-l^2Y(\theta),
\label{equ:2}
\end{equation}
where $\epsilon$ is the energy. The general solution of eq.2 being $Y(\theta)= a\exp(il\theta)+ b\exp(-il\theta)$, $l$ must be an integer to fulfil the periodic boundary conditions.

\begin{figure}
  \includegraphics[width=.95\linewidth]{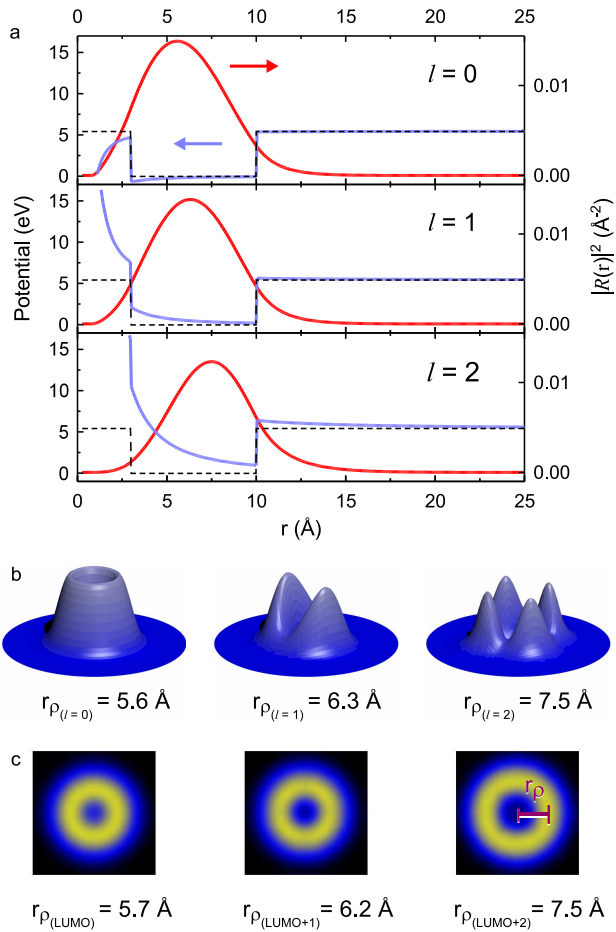}
  \caption{(Color online) (a) Finite square potential $V(r)$ (black dashed lines), effective potential $V_{eff}(r)$ (blue lines), and WGM probability density(red lines) for $l =$ 0, $l =$ 1 and $l =$ 2 modes. (b) 3D representations of the calculated WMG probability density  for the $l =$ 0, $l =$1, $l =$2  modes and for (a,b)=(0.25, 0.75). (c) Constant height differential conductance maps of the cyclic-12-thiophene (2.8 $\times$ 2.8 nm$^2$) at energies of the LUMO, LUMO+1 and LUMO+2. Experimental and calculated representations are at the same scale. $r_{\rho}$ is the radius defined as the distance from the center to the maxima of the circular resonances.} 
\label{fig4} 
\end{figure}

\noindent
Equation (\ref{eq:radial}) is the quantum expression of a wave equation whose solutions are 2D WGM.
The resemblance between a classical and a quantum representation of WGM has been highlighted recently for a spherical (3D) symmetry \cite{Pluchon2012}.  
For the 2D case, one defines the effective potential:
\begin{eqnarray*}
 V_{eff}(r)=V(r)+\frac{\hbar^2}{2m^{*}}\frac{1}{r^2}\left(l^2-\frac{1}{4}\right),
\end{eqnarray*}
where $V(r)$ depends on the nature of the wave (acoustic, optic, electronic...) and the geometry of the resonator, while $l^2-\frac{1}{4}$ is common to all type of waves and accounts for centrifugal forces (excepted for $l$ = 0 where the force is directed toward the center of the ring). Figure\,\ref{fig4}a represents the radial part of the states probability density $|{R(r)}|^{2}$ for $l = $ 0, 1 and 2 calculated from equation\,(\ref{eq:radial}) by using a Numerov procedure \cite{numerov1924}. For $l \neq$ 0, centrifugal forces modify the potential and produce a shift of the wavefunctions towards the outer wall with increasing $l$. This effect explains the increased radius of the nano-rings orbitals with energy observed in experimental conductance maps (Fig.\,\ref{fig2} i to k). For high values of $l$, a change of the inner wall radius has little influence on the position of the resonances, as expected for WGM \cite{rayleigh1904,Carrara1970,Hiremath2006}. Finaly, the $l = 0$ mode is localized to the inner wall. This surprising effect is linked to the 2D nature of the nanorings, and is therefore not observed in 3D resonators where the additional potential term is proportional to $l(l + 1)$.\\
Figure \ref{fig4}b represents the state probability density $|\psi(\vec{r})|^{2}$ in 3D for a given set of parameters (a,b).
These maps reveal the $\cos(2l\theta)$  modulated nature of waves for $l \neq$ 0 \cite{note2}, in apparent discrepancy with the uniform rings observed experimentally (\Figref{fig4}c). However, since there is
no privileged phase origin, the hamiltonian having no explicit angle dependence, all combinations of (a,b) are equally possible, leading to the uniform rings reported in the experimental maps. 

\noindent
In our model, the height of the potential (5.2 eV) and the effective electron mass (0.15 $m_e$) are deduced experimentally \cite{supmat}. The left and the right border of the potential (\Figref{fig4}a) are adjusted to reproduce the radial spread of the experimental orbitals (\Figref{fig4}c). An effective width of 7 \AA \,\,is estimated for polythiophene using this procedure. 
Figure \ref{fig3}b displays the dispersion curves of cyclic ($N$ = 12, 15, 30) wells for these parameters together with the limit case of a linear wire with infinite potential \cite{note3}. Qualitatively, the behaviour is the same than for the experiment: the dispersion of the ring's orbitals is larger than for the linear counterpart, and decreases as $N$ increases. Eventually, the dispersion of linear and cyclic wells merge for infinitely long wires. Quantitatively, the model overestimates the dispersions and the shapes of the curves are different. A more realistic description of the potential is likely to improve the agreement but would lack the simplicity of the square potential model.

\noindent
Despite the simple form of the potential, our 2D model reproduces the experimental observations with good accuracy and shows that oligothiophene nano-rings are resonators for electronic WGM. A fascinating consequence of this effect is the confinement of high-energy orbitals to the outer perimeter of the nano-rings, which also has important consequences in the scope of future applications: modes localized at the outer wall are more likely to couple adjacent cycles than modes confined to the inner wall. Thanks to this effect, a chain of cycles would act as a nanoscale electron energy high-pass filter. We speculate that molecular rings made of carbone nanotube slices \cite{Gleiter2009} would possibly fulfill this objective.    

\noindent
We thank F. Charra and L. Limot for stimulating discussions, and J.-G. Faullumel and M. Romeo for technical support. The Agence National  de  la  Recherche, contract TRANSMOL ANR-2010-JCJ-1004, the Région Alsace, and the International Center for Frontier Research in Chemistry (FRC) are acknowledged for financial  support.


\begin{thebibliography}{10}
\bibitem{note2}
The cases (a,b) = (0,1) and (1,0) constitute exceptions where $|Y(\theta)|^{2}$ is constant.
\bibitem{note3}
Note that for a linear wire with a finite potential, the wavefunctions no more vanish at the boundaries and they slightly spread out of the wire. This induces small deviations from the parabolic dispersion of the infinite potential, depending on the wire length and potential height.
\end{thebibliography}
\end{document}